\documentclass[10pt,aps,prl,superscriptaddress,twocolumn,preprintnumbers,nofootinbib]{revtex4-2}
\usepackage{amsmath,amssymb,mathtools}
\usepackage{booktabs}
\usepackage{graphicx}
\usepackage{lineno}
\usepackage{hyperref}
\newcommand{\sect}[1]{%
\bigskip\noindent%
{\bfseries\upshape\rmfamily\boldmath{#1}}---%
\ignorespaces%
}

\usepackage{color}
\definecolor{mygreen}{RGB}{0,128,0}

\definecolor{wfcolor}{RGB}{204,132,0}

\definecolor{myred}{RGB}{128,0,20}

\definecolor{nicered}{rgb}{0.7,0.1,0.1}
\definecolor{nicegreen}{rgb}{0.0,0.4,0.0}
\definecolor{niceblue}{rgb}{0.0,0.3,0.6}

\hypersetup{
    colorlinks=true,
    linkcolor=niceblue,
    citecolor=nicegreen,
    urlcolor=niceblue,
    }

\begin{document}

%\linenumbers

\preprint{CERN-TH-2025-018, IPPP/25/03, TUM-HEP-1553/25, ZU-TH 05/25}

\title{Precise Predictions for Event Shapes in Diphoton Production at the LHC}

\author{Federico Buccioni}
\email{federico.buccioni@tum.de}
 \affiliation{%
  Physics Department, Technical University Munich, James-Franck-Strasse 1, 85748 Garching, Germany}%
\author{Xuan Chen}
\email{xuan.chen@sdu.edu.cn}
 \affiliation{%
  School of Physics, Shandong University, Jinan, Shandong 250100, China}%
\author{Wei-Jie Feng}
\email{weijie.feng@physik.uzh.ch}
 \affiliation{%
  Physik-Institut, Universit\"at Z\"urich, Winterthurerstrasse 190, 8057 Z\"urich, Switzerland}%
\author{\\Thomas Gehrmann}
\email{thomas.gehrmann@uzh.ch}
 \affiliation{%
  Physik-Institut, Universit\"at Z\"urich, Winterthurerstrasse 190, 8057 Z\"urich, Switzerland}%
\author{Alexander Huss}
\email{alexander.huss@cern.ch}
\affiliation{%
Theoretical Physics Department, CERN, CH-1211 Geneva 23, Switzerland}%
\author{Matteo Marcoli}
\email{matteo.marcoli@durham.ac.uk}
\affiliation{%
  Institute for Particle Physics Phenomenology, Department of Physics, University of Durham, Durham, DH1 3LE, UK}%

\begin{abstract}
Photon pair production is an important benchmark process at the
LHC, entering Higgs boson studies and new physics searches. 
It has been measured to 
high accuracy, allowing for detailed studies of event shapes
in diphoton final states. To enable precision physics with
diphoton event shapes, we compute the second-order
QCD corrections, ${\cal O} (\alpha_s^3)$, to them and study their phenomenological impact. 
\end{abstract}

\maketitle

%---------------------------------------------------

\section{Introduction}

The production of photon pairs is a classical hadron 
collider observable. Following its initial observation at 
UA2~\cite{UA2:1992yma}, this process has been measured 
to increasing accuracy at the Tevatron~\cite{CDF:2012ool,D0:2013gxo} and the LHC~\cite{ATLAS:2021mbt,CMS:2014mvm}.
Photon pair final states played a crucial role in 
the discovery of the Higgs boson~\cite{ATLAS:2012yve,CMS:2012qbp} and in subsequent 
precision studies~\cite{ATLAS:2022fnp,CMS:2022wpo} of 
the Higgs boson properties. Photon pairs 
are also widely studied 
in searches for Physics beyond the Standard Model. 

The dominant Standard-Model production process for 
photon pairs is quark-antiquark annihilation. This
Born-level (leading order, LO) process receives large QCD corrections at 
next-to-leading order~\cite{Binoth:1999qq} 
(NLO, ${\cal O}(\alpha_s)$) and 
sizable ones at next-to-next-to-leading 
order~\cite{Catani:2011qz,Campbell:2016yrh,Gehrmann:2020oec} (NNLO, ${\cal O}(\alpha_s^2)$).
The Born-level process 
enforces the photons to be balanced in transverse momentum.
Therefore, the leading-order contribution 
to the diphoton transverse momentum distribution requires 
the presence of a partonic recoil in the final state, 
thus starting only at ${\cal O}(\alpha_s)$. The diphoton
transverse momentum distribution has also been 
computed~\cite{Chawdhry:2021hkp} to NNLO QCD, which in this case amounts to ${\cal O}(\alpha_s^3)$. Corresponding to 
a $2\to 3$ process at Born level, it is 
representative of the current frontier in computational 
complexity in NNLO QCD calculations. 

More detailed information on the production dynamics 
can be gained from the study of 
event shape distributions in diphoton final states. Event shapes describe geometrical properties of the final-state 
kinematics~\cite{Banfi:2010cf}. They take non-trivial values 
at Born level 
only for final states containing three or more objects. 
ATLAS have performed a detailed study~\cite{ATLAS:2021mbt} 
of diphoton final states at $\sqrt{s}=13$~TeV, measuring 
distributions in diphoton 
transverse momentum  and in three event shape 
variables: hadron collider thrust $a_T$, 
acoplanarity $\phi_{{\rm acop}}$, and decorrelation angle 
$\phi^*_\eta$. For events at low transverse momentum,
scaling relations between the transverse momentum and
each of these event shape variables can be 
established~\cite{Banfi:2010cf}. These demonstrate that the 
event shape distributions are measured with considerably higher 
resolution than the transverse momentum distribution, mainly 
owing to the fact that the event shapes rely on the 
determination of the directions of the photons and not on 
their energies, which are harder to resolve experimentally.

The event shape distributions are thus ideally suited 
for high-resolution studies of the QCD dynamics in 
the production of photon pairs. In the case of the ATLAS
measurement~\cite{ATLAS:2021mbt}, 
a complex interplay between the fiducial 
selection cuts, applied on the individual photons, 
and the event shape definition on the photon pair momenta
is taking place. These effects can often only be resolved 
by taking into account higher-order QCD corrections, 
especially from
real radiation and recoils. 
In this letter, we compute the NNLO QCD corrections,
${\cal O} (\alpha_s^3)$, to diphoton event shape 
distributions and perform detailed studies of their 
phenomenological impact.

\section{Methodology}
As for the transverse momentum distribution,
non-trivial contributions to the diphoton 
event shape distributions
are generated only in the presence of a partonic recoil. The 
underlying QCD process is therefore 
\begin{equation}
	pp\to \gamma \gamma +{\rm jet} + X,\nonumber
\end{equation}
with the jet definition replaced by a minimum cut on an event 
shape variable or on the diphoton transverse momentum.
In order to single out direct-photon production and remove
hadronic contamination, photons are identified by an isolation 
criterion defined through a cone around the photon direction. 
NLO calculations~\cite{Binoth:1999qq} mirror the 
exact isolation prescription that is used in the experimental 
measurements. At NNLO, an idealized photon isolation 
(dynamical cone~\cite{Frixione:1998jh} or hybrid 
cone~\cite{Siegert:2016bre}) is used.  

Renormalized one- and two-loop corrections to the Born process contain explicit infrared (IR) singularities arising from loop-momentum integration, whereas real-emission matrix elements for the radiation of one or two additional partons exhibit a divergent behavior in soft and collinear configurations. Upon combination of virtual and real corrections,
infrared divergences cancel in theoretical predictions for physical observables, but dedicated techniques have to be employed to achieve such cancellation. Our calculation uses the well-established antenna subtraction method~\cite{Gehrmann-DeRidder:2005btv,Daleo:2006xa,Currie:2013vh},
which extracts the 
divergent behavior of real-emission corrections locally across phase space by suitable subtraction terms constructed using \textit{antenna functions}. The analytically integrated counterpart of such terms is then used to cancel the explicit singularities of virtual corrections. The numerical integration of infrared-finite remainders is performed within the Monte Carlo event generator \textsc{NNLOjet}~\cite{Gehrmann-DeRidder:2015wbt,NNLOJET:2025rno}. 

The six-point one-loop and seven-point tree-level matrix elements for the real-virtual and double-real corrections are obtained from 
\textsc{OpenLoops2}~\cite{Cascioli:2011va,Buccioni:2019sur}. 
The two-loop amplitudes are 
expressed through their IR-subtracted remainder functions~\cite{Catani:1998bh}. They have initially been computed in the leading-color approximation~\cite{Agarwal:2021grm,Chawdhry:2021mkw} (which was used 
in~\cite{Chawdhry:2021hkp})
and later in full color~\cite{Agarwal:2021vdh}.
The two-loop amplitudes are expressed
as linear combinations of massless two-loop five-point integrals~\cite{Chicherin:2018old}, which are 
evaluated in terms of so-called pentagon functions~\cite{Chicherin:2020oor}. For the first time, 
our calculation includes these amplitudes in full color, allowing us to quantify 
the impact of subleading-color effects.

Due to the large gluon luminosity at the LHC, diphoton cross sections receive sizable contributions
from loop-induced gluon-initiated reactions, where the photon pair is radiated off a closed quark loop. 
The Born level contribution to this process 
 corresponds to the square of 
a one-loop amplitude, which 
thus starts contributing 
only at $\mathcal{O}(\alpha_s^3)$, as part of 
the full NNLO contribution. 
Owing to this delayed perturbative onset, 
the loop-induced gluon fusion contribution 
has been identified to be responsible for a substantial fraction of 
the scale uncertainty in diphoton-plus-jet 
cross sections at NNLO~\cite{Chawdhry:2021hkp}. 
Higher-order corrections to this process, originally studied in~\cite{Badger:2021ohm} for the gluon-initiated channels only, are formally beyond NNLO accuracy, but they can reduce the theoretical uncertainty. 
In this letter, we assess the impact of the complete 
NLO, $\mathcal{O}(\alpha_s^4)$,
correction to this loop-induced process, including all its partonic initial states. 
The virtual corrections to this process are given by the interference of five-point one- and two-loop amplitudes. The full-color two-loop contributions of the $gg\to\gamma\gamma g$ process were presented in~\cite{Badger:2021imn} and are distributed through the \textsc{NJet} library~\cite{Badger:2012pg}. The analogous amplitudes for the $q\bar{q}\to\gamma\gamma g$ process, and crossings thereof, were computed in~\cite{Agarwal:2021vdh}. Real-radiation corrections, which entail six-point one-loop squared amplitudes, are obtained from \textsc{OpenLoops2} in all partonic channels.

The \textsc{NNLOjet} implementation was validated in various ways. For the real corrections, the numerical evaluation of antenna subtraction terms was tested pointwise against \textsc{OpenLoops2} matrix elements in each IR-singular phase-space region. For the virtual ones, 
we checked that the universal IR-pole structure of one- and two-loop amplitudes~\cite{Catani:1998bh}
is reproduced by the integrated antenna functions. At cross-section level, we compared our results for
diphoton plus one and two jets at NLO accuracy against \texttt{MadGraph5\_aMC@NLO}~\cite{Alwall:2014hca,Frederix:2018nkq}, finding agreement.
Finally, at NNLO we reproduce the 
diphoton-plus-jet differential cross sections 
of~\cite{Chawdhry:2021hkp},  
by truncating the  
two-loop virtual corrections (VV) to leading-color
for consistency.

\section{Results}

The ATLAS 13~TeV analysis \cite{ATLAS:2021mbt} of diphoton event shapes is performed in the fiducial region defined by the following cuts:
\begin{eqnarray*}
 && p_{T,\gamma_{1}} > 40~{\rm{GeV}}, \quad   p_{T,\gamma_{2}} > 30~\rm{GeV}, \\
 && |\eta_\gamma|\in(0,1.37)\cup(1.52,2.37), \quad 
 \Delta R_{\gamma\gamma}>0.4\,.
\end{eqnarray*}
The photons are identified by a fixed-cone isolation criterion,
allowing a maximum hadronic energy fraction $\epsilon_{T,\gamma}$
inside a cone of radius $R$ around the photon direction, 
with $(R,\epsilon_{T,\gamma}) = (0.2,0.09)$. We note that this $R$ value
is the same as used in Higgs-to-diphoton studies~\cite{ATLAS:2022fnp}, but 
is lower than $R=0.4$ that is typically used 
in single-photon measurements~\cite{ATLAS:2023yrt}.

In our theoretical prediction we use the Parton Distribution Function (PDF) set \texttt{PDF4LHC21\_mc}~\cite{PDF4LHCWorkingGroup:2022cjn} via the \texttt{LHAPDF6}~\cite{Buckley:2014ana} interface and we adopt a hybrid-cone isolation prescription~\cite{Siegert:2016bre} which inserts a dynamical cone~\cite{Frixione:1998jh} of parameters $(R_d, \epsilon_d, n)=(0.1, 0.15, 1)$ inside the fixed isolation cone. 
The renormalization and factorization scales are chosen dynamically using the central scale $\mu_0^2=E_{T,\gamma\gamma}^2=m_{\gamma\gamma}^2 + p_{T,\gamma\gamma}^2$. Theoretical uncertainties are estimated via 7-point scale variations, namely by varying $\mu_R$ and $\mu_F$ with multiplicative factors $\xi_{R,F}\in[1/2,2]$ imposing $1/2\leqslant \mu_R/\mu_F \leqslant 2$.

In total, the ATLAS measurement presents one-dimensional 
distributions in eight different diphoton variables. Four of 
these distributions, $p_{T,\gamma_{1}}$, $p_{T,\gamma_{2}}$,
$m_{\gamma\gamma}$, and $|\cos \theta^*|_{{\rm CS}}$,
receive Born-level contributions already at ${\cal O}(\alpha_s^0)$. They are compared to NNLO QCD 
predictions~\cite{Gehrmann:2020oec} from \textsc{NNLOjet}
already in~\cite{ATLAS:2021mbt}. The four remaining 
distributions start only at $\mathcal{O}(\alpha_s)$: transverse 
momentum of the diphoton system $p_{T,\gamma\gamma}$ and 
the three event shapes~\cite{Banfi:2010cf}:
\begin{eqnarray}
a_T &=& 2\frac{|p_{x,\gamma_1}p_{y,\gamma_2} 
- p_{y,\gamma_1}p_{x,\gamma_2} |} 
{|\vec{p}_{T,\gamma_1} - \vec{p}_{T,\gamma_2}|}\,, \\
\phi^*_\eta &=& \tan\frac{\pi-\Delta \phi_{\gamma\gamma}}{2}
\sqrt{1-\tanh^2(\Delta\eta_{\gamma\gamma}/2)}\,, \\
\phi_{{\rm acop}}&=&\pi-\Delta\phi_{\gamma\gamma}\,.
\end{eqnarray}
Our NNLO QCD calculation enables accurate predictions for these four observables,
and in the following we exclusively focus on them.

\begin{table}[t]
    \centering
\begin{tabular}{llc}
  \toprule
  Observable & Cut &
 Equivalent $p_{T,\gamma\gamma}$ [GeV] \\
  \midrule
  $p_{T,\gamma\gamma}$ [GeV]  & 1.0 & $-$\\
  $a_{T}$ [GeV] & 1.08 & 1.52 \\
  $\phi^*_\eta$ & 0.0105 & 1.19 \\
  $\phi_{\rm{acop}}$ & 0.0234 & 1.12 \\
  \bottomrule
\end{tabular}
\caption{
Lower cuts on the event shapes
applied in our comparison to the ATLAS data~\cite{ATLAS:2021mbt} and 
their conversion to an equivalent value of 
$p_{T,\gamma\gamma}$, assuming  
$m_{\gamma\gamma}=80$~GeV.}
\label{tab:obscutsbrk}
\end{table}
At the lower endpoint of event shape distributions, 
the fixed-order description breaks down due to the emergence 
of large logarithmic corrections at each order in perturbation 
theory that need to be resummed for reliable predictions. 
Also, in these endpoint regions the numerical evaluation of the fixed-order
components becomes increasingly challenging. We therefore cut off the 
distributions 
at bin edges in the respective measurement, 
listed in Table~\ref{tab:obscutsbrk}.
%\begin{eqnarray}
%    &&p_{T,\gamma\gamma} > 1~\mbox{GeV}\,, \quad
%    a_{T} > 1.08~\mbox{GeV}\,, \nonumber \\
%    &&\phi^*_\eta > 0.0105\,, \quad \phi_{{\rm acop}}>0.0234\,,
%    \label{eq:lowcut}
%\end{eqnarray}
%each corresponding to bin edges in the respective measurement. 
At the event-generation level, an event is accepted if it passes at least one of these cuts, which
ensures the infrared-safety of the evaluation. 
\begin{figure*}[t]
 \centering
\hspace*{-0.5cm}\includegraphics[width=0.48\textwidth]{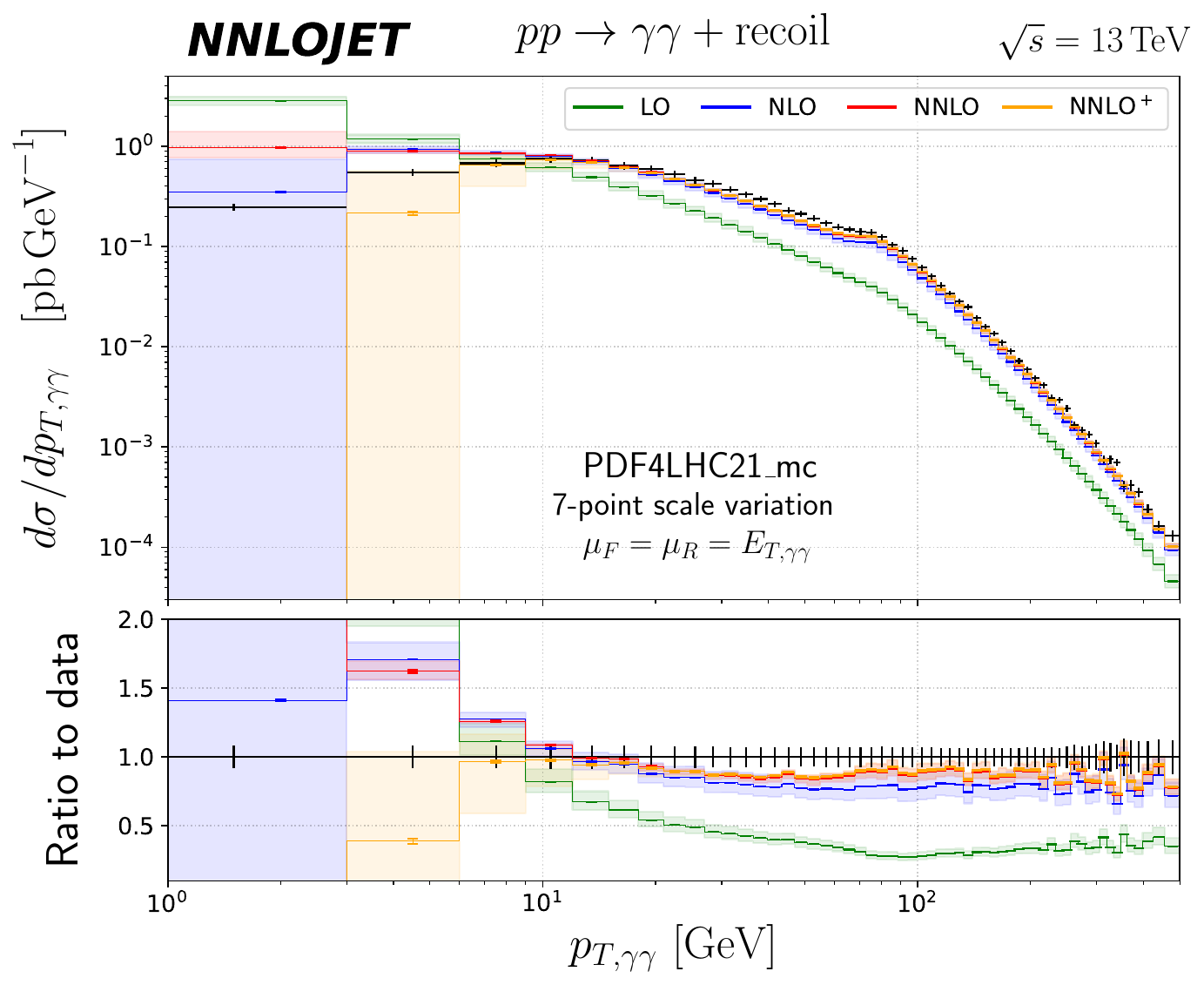}
\hspace*{0.2cm}\includegraphics[width=0.48\textwidth]{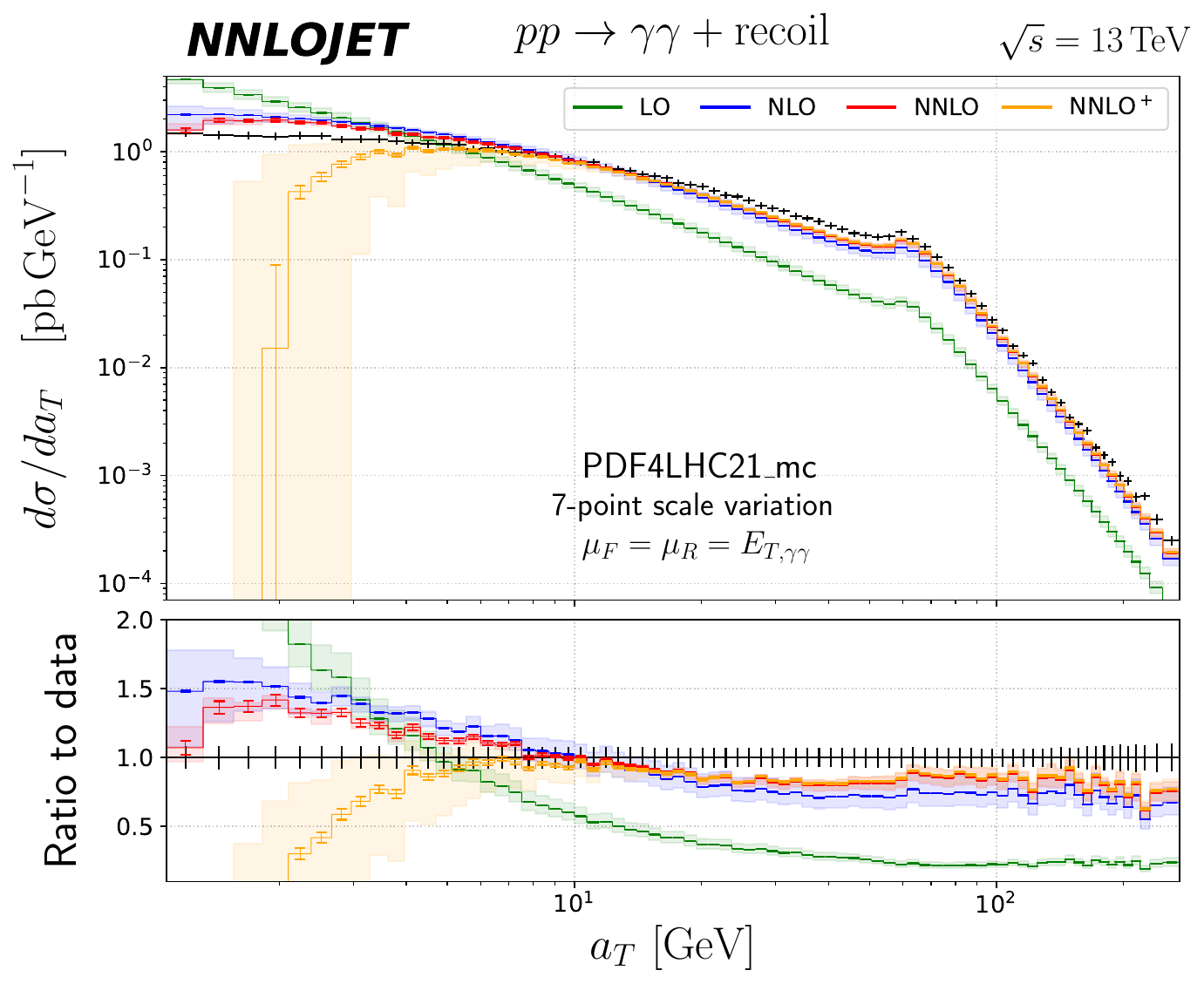}\\
\hspace*{-0.5cm}\includegraphics[width=0.48\textwidth]{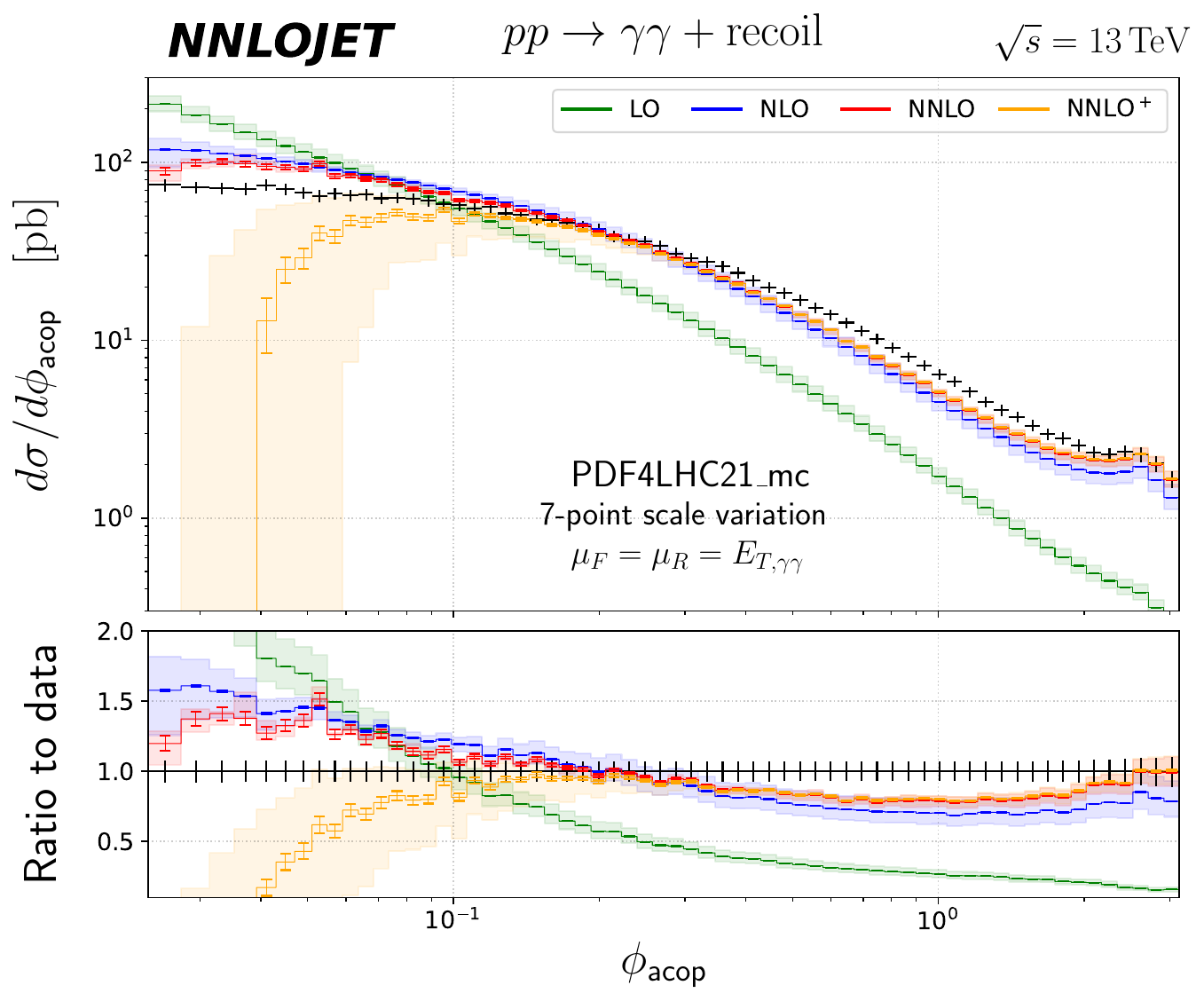} 
\hspace*{0.2cm}\includegraphics[width=0.48\textwidth]{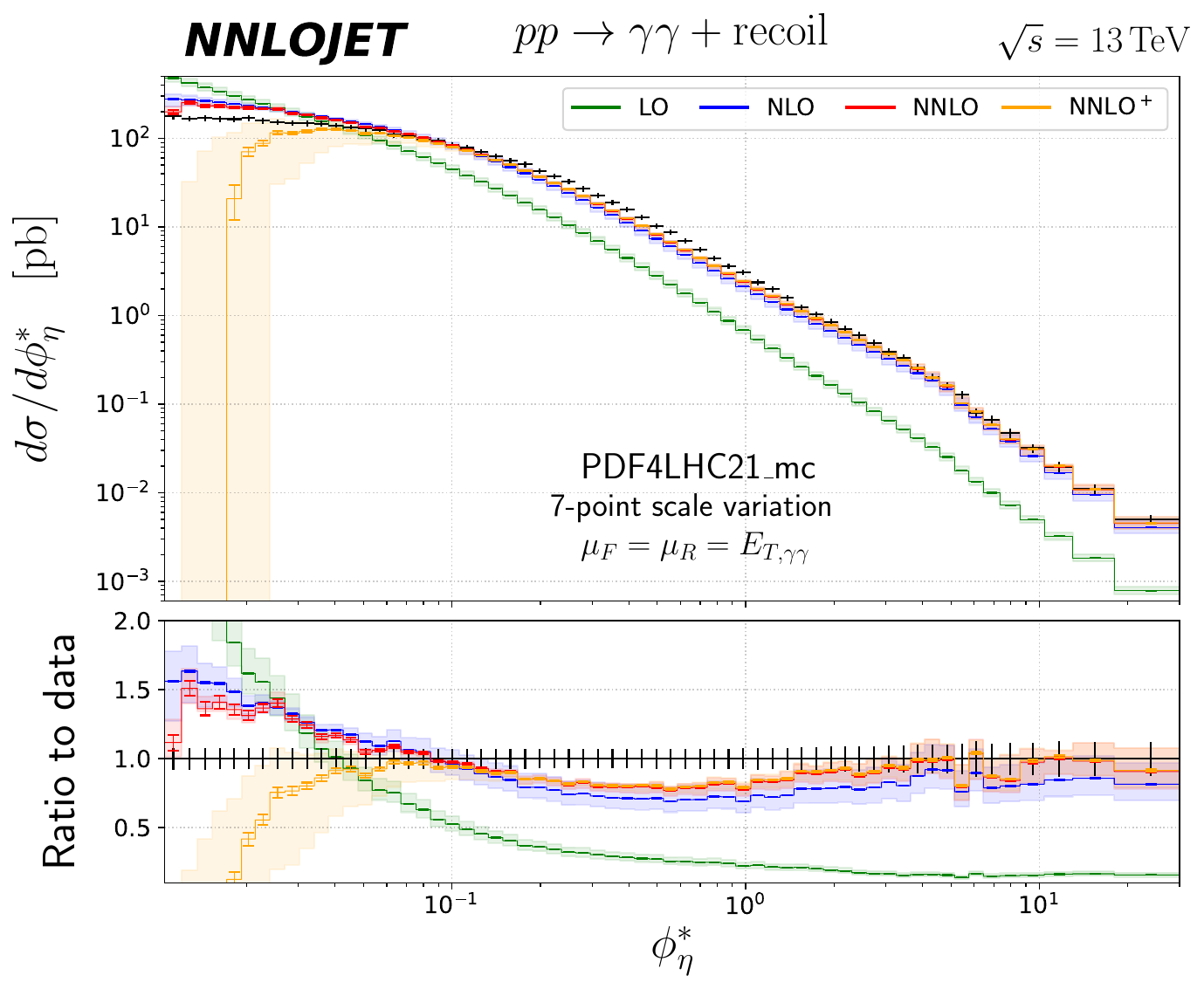}
\caption{$p_{T,\gamma\gamma}$, $a_T$, $\phi_{\text{acop}}$ and 
$\phi^{*}_\eta$ distributions for diphoton production at the LHC in LO (green), NLO (blue) and NNLO (red) accuracy, in comparison with the ATLAS diphoton measurement~\cite{ATLAS:2021mbt}. The NNLO prediction consistently includes the loop-induced gluon-initiated process at Born level. The NNLO$^+$ curve (yellow) is obtained by
including the $\mathcal{O}(\alpha_s^4)$ NLO correction to the loop-induced process. The colored bands represent theoretical uncertainties from 7-point scale variation. The error bars represent Monte-Carlo integration errors. The ratio plots show the prediction normalized to the ATLAS data.}
\label{fig:fullNNLOvsATLAS}
\end{figure*}

Towards lower values of transverse momentum
or event shapes, the final-state photons approach a back-to-back configuration,
leading to a diphoton-invariant mass
of $m_{\gamma\gamma}\geqslant 80$~GeV, corresponding to
twice the cut on $p_{T,\gamma_{1}}$. 
In these back-to-back
configurations, 
the cutoff values in Table~\ref{tab:obscutsbrk}
are related by the approximate scaling~\cite{Banfi:2010cf,Gehrmann-DeRidder:2016jns}:
\begin{eqnarray}
p_{T,\gamma\gamma} &\approx& \sqrt{2} a_T \,, 
\nonumber \\
p_{T,\gamma\gamma}/m_{\gamma\gamma}
&\approx& \sqrt{2} \phi^*_\eta
\;\approx\; 0.85\sqrt{2}
\tan(\phi_{\rm{acop}}/2)\,.
\label{eq:approx}
\end{eqnarray}

Figure~\ref{fig:fullNNLOvsATLAS} presents the theoretical prediction up to NNLO in comparison to 
the ATLAS data~\cite{ATLAS:2021mbt}. Overall, we
observe that the inclusion of NNLO corrections 
leads to an improved description of the data 
in the region where the transverse momentum 
or the values of the event shape 
variables are sufficiently large. In these regions, 
the NNLO corrections are positive and the 
scale uncertainty on the theory predictions typically 
drops from around $\pm(10\ldots15)$\% at NLO to $\pm(2\ldots10)$\%  at NNLO.
Although the NNLO corrections bring the NLO curve closer to the data in these regions, 
the theory predictions still fall systematically below the measurements. The ATLAS study~\cite{ATLAS:2021mbt} indicates the potential relevance of
hadronization effects in these regions, which are
not accounted for in our parton-level calculation. 

Towards the lower end of all distributions
(below $p_{T,\gamma\gamma}\approx 10$~GeV and 
equivalent values of the shape variables), 
the NLO and NNLO predictions are 
systematically above the experimental data. 
NNLO corrections are still moderate in the 
kinematical range displayed in 
Figure~\ref{fig:fullNNLOvsATLAS}, and 
an apparent perturbative convergence is observed
in this range, with NNLO predictions typically 
within the NLO scale uncertainty, even if  
here 
one expects all-order resummation to be needed for reliable predictions.
We observe in particular that the bin resolution in this region is 
considerably higher for  
 $a_{T}$,  $\phi^*_\eta$ and $\phi_{{\rm acop}}$ than it is for 
 $p_{T,\gamma\gamma}$. This superior resolution allows to probe this infrared-sensitive 
 region
 considerably more accurately than through the $p_{T,\gamma\gamma}$ spectrum.

Above $p_{T,\gamma\gamma} \approx 70$~GeV, 
we observe a bump in the transverse momentum 
distribution, which is a consequence of the
the fiducial cuts that are applied in 
the measurement.
No explicit 
cut is imposed on the diphoton-invariant mass,
which is only implicitly restricted from below 
through the photon 
angular separation $\Delta R_{\gamma\gamma}>0.4$. 
Low invariant-mass photon pairs require both photons 
to be in the same hemisphere, such that their combined transverse momentum must be balanced by a partonic 
recoil. Consequently, these low-mass 
photon pairs contribute 
to the transverse momentum distribution only 
above $p_{T,\gamma\gamma} \sim 70$~GeV, corresponding to the sum of the individual $p_{T,\gamma}$ cuts. 
Similar features, though less pronounced, are observed in 
all event shape distributions. Also, they are
stable under perturbative corrections, 
indicating that they do not give rise to large logarithms. 

The NLO, $\mathcal{O}(\alpha_s^4)$, correction to the process where both photons couple to a 
closed quark loop is added to the 
NNLO, $\mathcal{O}(\alpha_s^3)$, predictions to 
yield the yellow NNLO$^+$ curves in 
Figure~\ref{fig:fullNNLOvsATLAS}. At large
values of event shapes and 
$p_{T,\gamma\gamma}$, the inclusion of this 
contribution does not reduce the 
 NNLO scale uncertainty in a visible manner,
 contrary to initial expectations~\cite{Chawdhry:2021hkp,Badger:2021ohm}. 
For medium and low values of event shapes 
and transverse momentum, the NLO corrections 
to this subprocess become large and negative, 
leading to a substantial deterioration of 
the perturbative convergence of the NNLO$^+$
predictions. 
 \begin{figure*}[t!]
 \centering
\hspace*{-0.5cm}\includegraphics[width=0.46\textwidth]{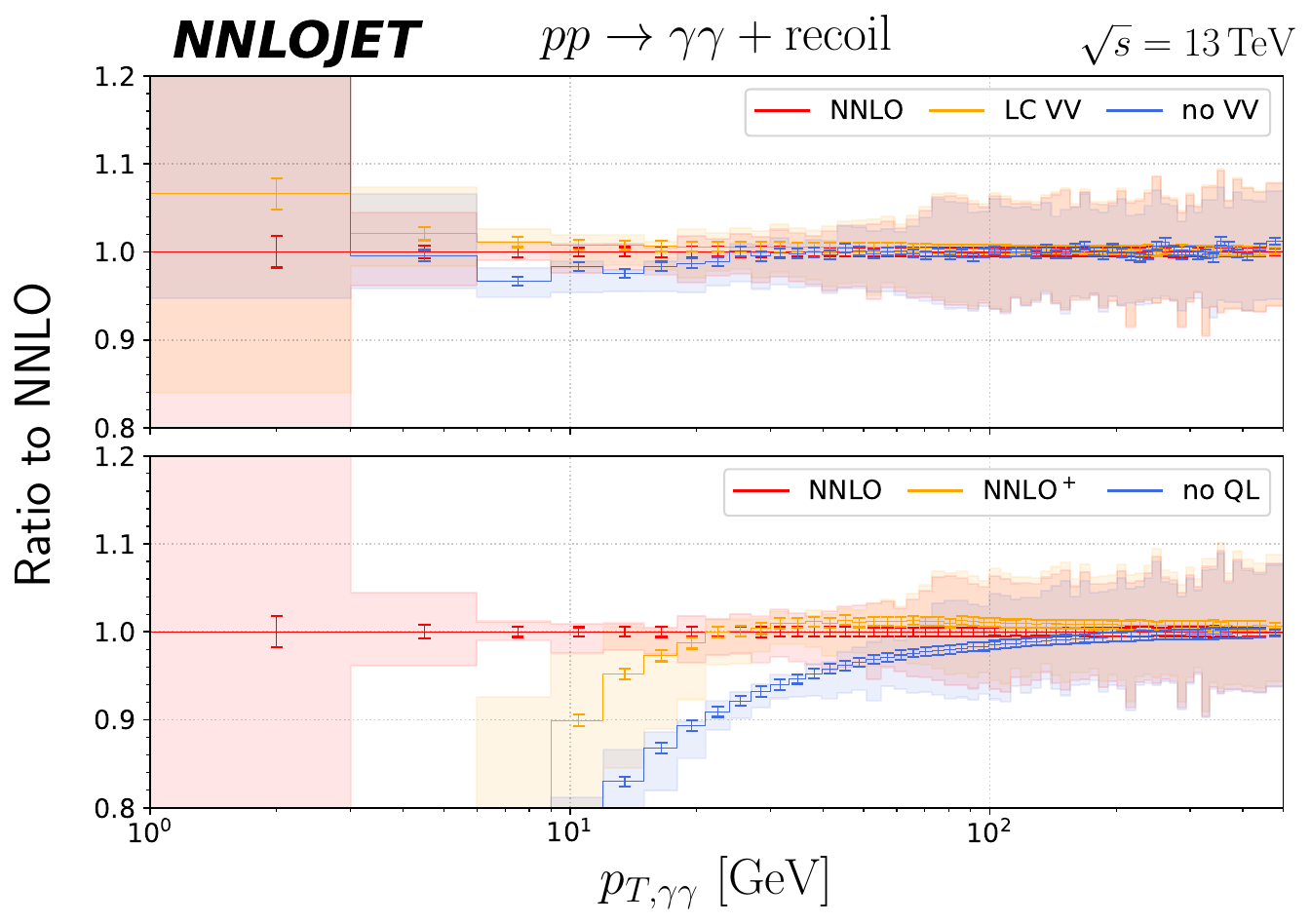}\,
\hspace*{0.3cm}\includegraphics[width=0.46\textwidth]{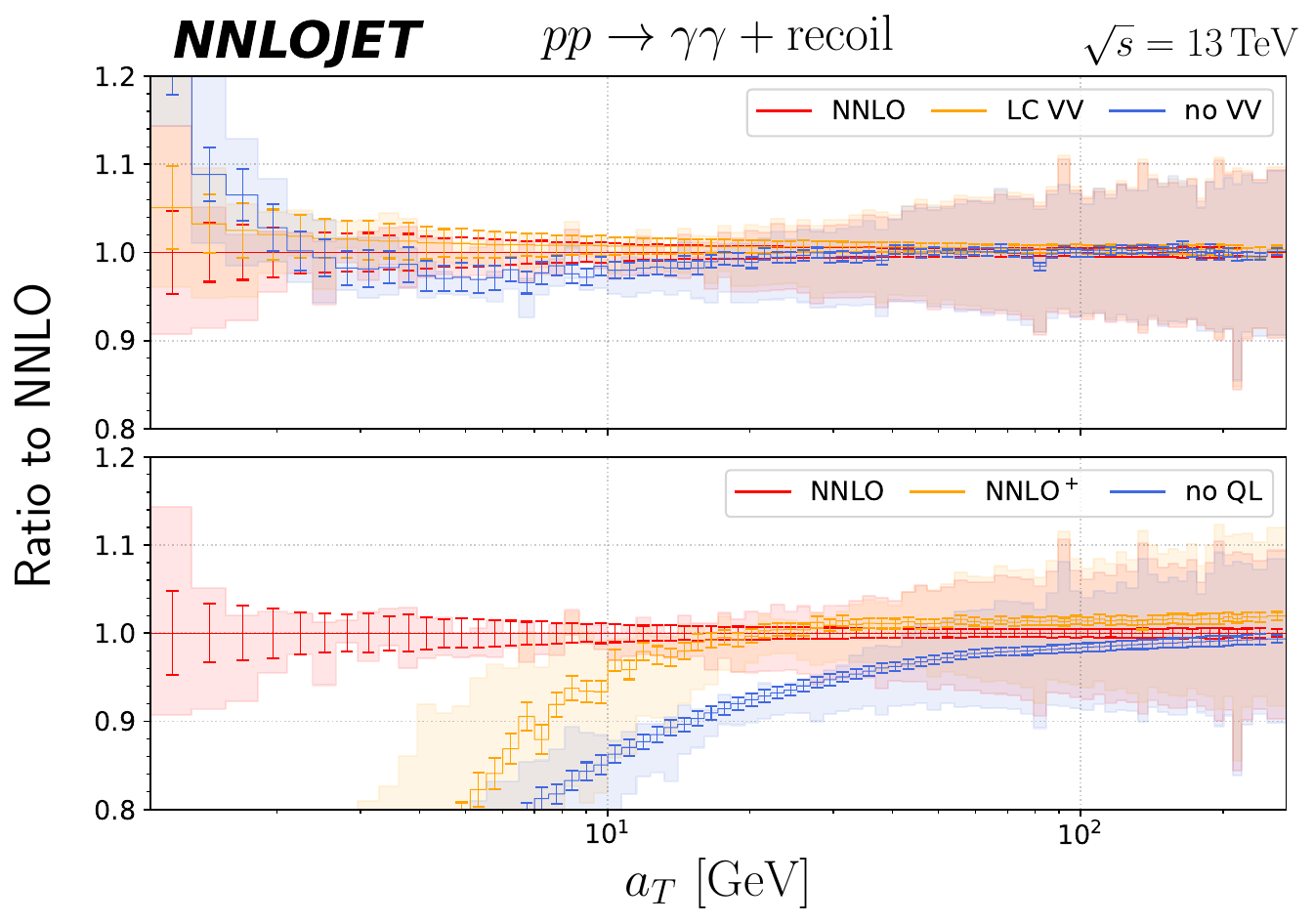}
\\
\hspace*{-0.5cm}\includegraphics[width=0.46\textwidth]{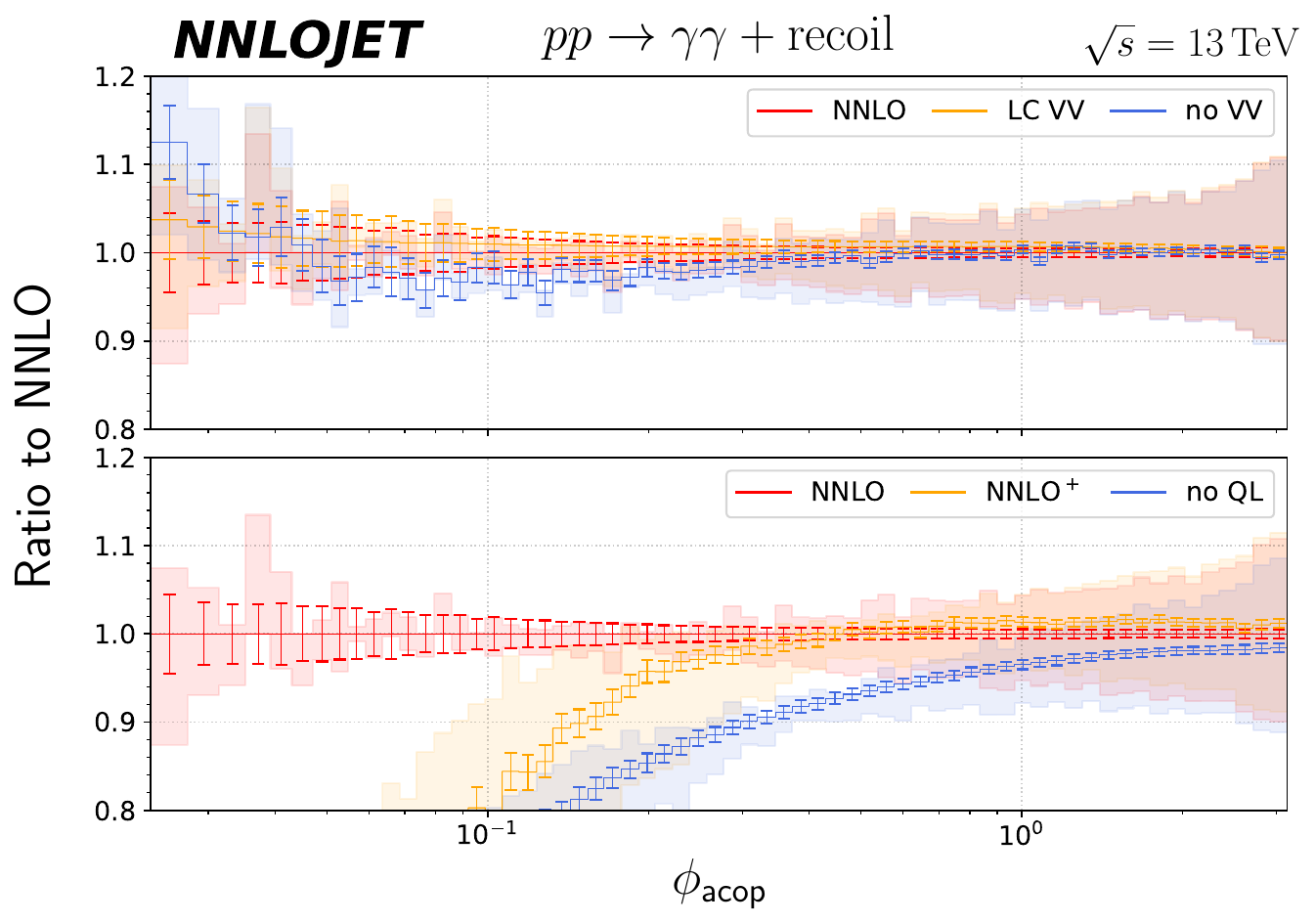}\,
\hspace*{0.3cm}\includegraphics[width=0.46\textwidth]{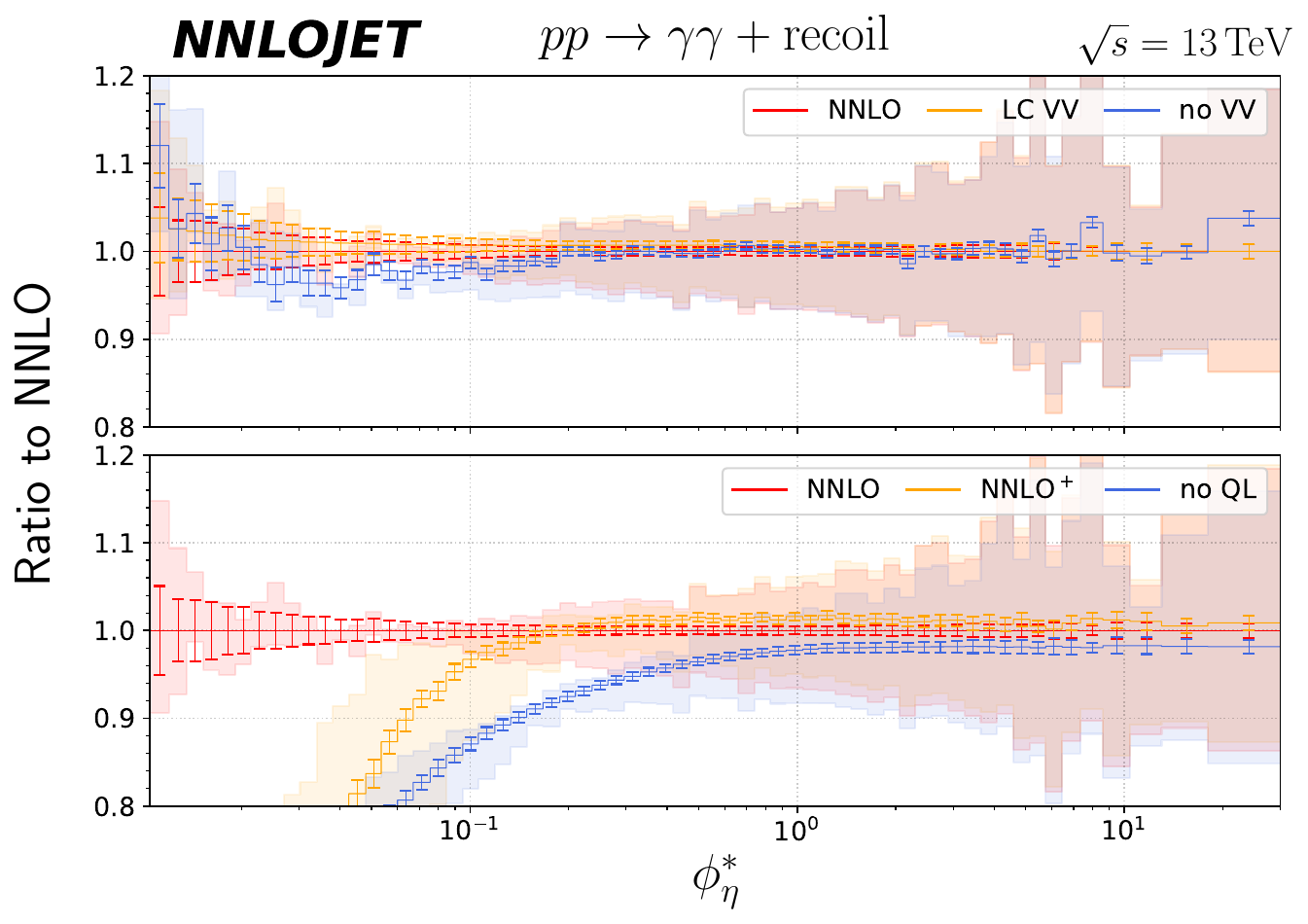}
\caption{ Numerical impact of virtual two-loop corrections (upper frames): comparison of the full calculation at NNLO (red, `NNLO'), calculation without the two-loop finite remainder $2\text{Re}\left(A_0 A_2^*\right)$ (blue, `no VV') and  calculation obtained including only the leading-color contribution to the two-loop finite remainder (yellow, `LC VV'). Numerical impact of loop-induced 
 processes (lower frames): comparison of the full calculation at NNLO (red, `NNLO'), calculation without the gluon-initiated $\mathcal{O}(\alpha_s^3)$ 
 quark-loop contribution (blue, `no QL') 
 and calculation obtained including it and the $\mathcal{O}(\alpha_s^4)$ NLO correction (yellow, `NNLO$^+$'). } 
\label{fig:NNLOggEffects}
\end{figure*}

Compared to previous NNLO results for diphoton final states at finite transverse momentum~\cite{Chawdhry:2021hkp}, we now take into account
the complete color information at two-loop level.
The 
newly included subleading color terms receive contributions from non-planar two-loop five-point amplitudes, which are also present in the 
the gluon-induced quark-loop subprocess at 
order $\alpha_s^4$.  
We quantify the numerical impact of these
newly included terms in Figure~\ref{fig:NNLOggEffects}, which displays the 
$p_{T,\gamma\gamma}$, $a_{T}$,  $\phi^*_\eta$ and $\phi_{{\rm acop}}$ distributions, 
comparing the full NNLO with the results obtained by neglecting or including specific contributions. 

The impact of the finite remainder of the 
two-loop virtual corrections is indicated in the upper frames of Figure~\ref{fig:NNLOggEffects}. More specifically, we isolate the interference of the two-loop amplitude with the tree-level one ($2\text{Re}(A_0 A_2^{*})$) and we either remove it completely (label `no VV') or we include only its leading-color component (label `LC VV'). In terms of two-loop Feynman diagrams, the leading-color component is defined as in~\cite{Chawdhry:2021hkp}. 
However, we note that our definition of finite remainder~\cite{Catani:1998bh,Agarwal:2021vdh} differs from the one of~\cite{Chawdhry:2021mkw,Chawdhry:2021hkp}. 
It can be seen that the contribution of the complete two-loop finite remainder depends non-trivially on the kinematics. It is quasi negligible for large values of the event 
shapes or $p_{T,\gamma\gamma}$, and typically 
amounts to less than 5\% of the NNLO prediction
except at the low endpoints of the 
distributions.  This is largely dominated by the leading-color component, with subleading color terms never exceeding the 0.3\% level.

In the lower frame of Figure~\ref{fig:NNLOggEffects}, we study the impact of the gluon-induced quark-loop subprocess, 
which is by default included at 
order $\alpha_s^3$ in the NNLO predictions. Its numerical 
relevance can be seen from the `no QL', where 
this subprocess is removed entirely. In the 
NNLO$^+$ curves, the $\mathcal{O}(\alpha_s^4)$ corrections
to this subprocess are included. 
It can be seen that 
the quark-loop subprocess contributes 
in particular in the low and intermediate range 
in the distributions. Its inclusion at NNLO 
is crucial for the perturbative stability of 
the predictions at this order, 
especially towards the lower 
end of the distributions. This feature is remarkable, since
at NNLO this contribution appears for the first
time, being 
finite and unrelated to any of the other subprocesses. 
The $\mathcal{O}(\alpha_s^4)$ corrections to the quark-loop 
subprocesses, included in NNLO$^+$,
are very small in the bulk of the distributions, and become very sizable and negative only 
towards their lower ends. Their smallness in the bulk explains their negligible impact on the scale uncertainty, where a small improvement is 
observed only in the medium range of the distributions. This effect quickly deteriorates 
at lower values. 

\section{Conclusions}
In this letter, we computed the NNLO QCD corrections to the diphoton transverse momentum distribution and 
to event shape distributions related to it. The corrections considerably improve the description of ATLAS 
data~\cite{ATLAS:2021mbt} for these precision observables. The distributions display several kinematical features, 
which can be explained through an intricate interplay between the diphoton observables definition and the fiducial cuts on the individual photons.  We quantified the numerical impact of previously unaccounted contributions from process classes containing two-loop non-planar 
virtual corrections, supporting the validity of leading-color truncations for generic two-loop amplitudes. We investigated the 
impact of quark-loop induced subprocesses, 
demonstrating their relevance at low and 
intermediate values of the shape variables at 
NNLO, and highlighting the need for their 
consistent inclusion at order $\alpha_s^3$.  

 Our results will enable precision phenomenology studies with 
event shape distributions in
diphoton final states. By demonstrating the 
numerical convergence and stability of the 
NNLO predictions at low values of transverse momentum, they also represent an important step towards the third-order QCD 
corrections for more inclusive diphoton observables.

\begin{acknowledgments}
\sect{Acknowledgments}
The authors would like to thank HanTian Zhang and 
Florian Lorkowski for useful discussions and for help with 
the usage of the external libraries. 
This work has received funding from the Swiss National Science Foundation (SNF)
under contract 200020-204200 and from the European Research Council (ERC) under
the European Union's Horizon 2020 research and innovation program grant
agreements 101019620 (ERC Advanced Grant TOPUP) and 949279 (ERC Starting Grant \textsc{HighPHun}), and from the National Science Foundation of China (grant No.12475085 and No.12321005). MM is supported by a Royal Society Newton International Fellowship (NIF/R1/232539).
\end{acknowledgments}

\bibliography{gagajnnlo}

\end{document}